\documentclass[conference, 10 pt]{IEEEtran}  

\IEEEoverridecommandlockouts

\usepackage{cite}
\usepackage{amsmath,amssymb,amsfonts}
\usepackage{algorithmic}
\usepackage{graphicx}
\usepackage{subfig}
\usepackage{orcidlink}
\usepackage{booktabs}
\usepackage{multirow}
\usepackage[normalem]{ulem}
\usepackage[utf8]{inputenc} 
\usepackage{booktabs}

\usepackage{booktabs}
\usepackage{threeparttable}
\usepackage{siunitx}
\usepackage{multirow}
\usepackage[T1]{fontenc} 
\usepackage{textcomp}
\usepackage{xcolor}
\def\BibTeX{{\rm B\kern-.05em{\sc i\kern-.025em b}\kern-.08em
    T\kern-.1667em\lower.7ex\hbox{E}\kern-.125emX}}

\usepackage{geometry}
\usepackage{float}

\begin{document}


\title{In-Vehicle Human–Machine Interface to Support Drivers in Conditionally Automated Platooning}

\author{
Anna-Lena~Hager\textsuperscript{1}\,\orcidlink{0009-0001-9738-3444}, 
\textit{Graduate Student Member, IEEE}, 
Mohamed~Sabry\textsuperscript{1}\,\orcidlink{0000-0002-9721-6291},
\textit{Graduate Student Member, IEEE}, \\
Walter~Morales-Alvarez\textsuperscript{1}\,\orcidlink{0000-0001-6912-4130},
\textit{Graduate Student Member, IEEE}, Selena~M\"ohrlein, and \\
Cristina~Olaverri-Monreal\textsuperscript{1}\,\orcidlink{0000-0002-5211-3598},
\textit{Senior Member, IEEE}
\thanks{ \textsuperscript{1}Dept.  Intelligent Transport Systems, Johannes Kepler University Linz, Altenberger Straße~69, 4040~Linz, Austria. 
\texttt{\{anna-lena.hager, mohamed.sabry, walter.morales\_alvarez, cristina.olaverri-monreal\}@jku.at},
\texttt{selenamoehrlein@gmx.de} }%
}
\maketitle
\begin{abstract}
Vehicle platooning enables close-gap driving and offers potential benefits for traffic efficiency and safety. In conditionally automated platooning, drivers remain responsible for supervising the system and intervening when necessary, making effective Human–Machine Interfaces (HMIs) critical for maintaining situational awareness and stable driver–automation coordination. This paper investigates whether an in-vehicle HMI providing continuous system-state and inter-vehicle distance information improves supervisory behavior, safety, and platoon stability. We conducted a simulation-based experiment integrated with a 6-degree-of-freedom motion system to enhance scenario realism. Dependent variables included collision occurrence, response latency following platoon disconnection, and the number of manual interventions during intact platooning.

Results showed significantly fewer manual interventions when the HMI was active, with intervention rates about 80\% higher without it. No significant effects were found for collision occurrence or response latency, indicating that additional information improves supervisory stability during platooning but does not substantially affect emergency reactions or collision rates.


\end{abstract}


\section{Introduction}
Platooning is the coordination of vehicles that follow each other closely to enhance road capacity and traffic flow, while potentially improving highway safety and reducing both energy consumption and emissions~\cite{campolo_vehicular_2015}. 
It can be realized through various technical approaches, including Adaptive Cruise Control (ACC) and Cooperative Adaptive Cruise Control (CACC). 
ACC uses range sensors to regulate vehicle speed and maintain a desired headway to the lead vehicle~\cite{ulsoy_automotive_2012}, while CACC extends this capability by incorporating vehicle-to-vehicle communication to exchange motion information ~\cite{van2006impact}. 
The availability of additional information enables tighter vehicle spacing and improved platoon stability. In situations where such information is unavailable, alternative cruise control strategies with minimal data requirements can be employed as fallback solutions, such as the Fallback Longitudinal Controller presented in ~\cite{sabry2025lidar}.

Through the use of these technologies, parts of the platooning task are automated, resulting in a shift of control authority from the driver to the system. As a result, maintaining awareness of the automation status information is essential \cite{sarter_how_1995}. 
When a system reaches its Operational Driving Domain (ODD) limits or failures arise, requiring driver intervention, accessing control-relevant system state information is crucial to enable timely and accurate decision-making in such events \cite{jamieson_designing_2005}. 
In this paper, we developed a Human–Machine Interface (HMI) for close-interval platooning scenarios that provides drivers with additional information about the platoon in operation, such as the distance to the lead vehicle and the platoon's status. 
The scenarios were implemented on a 6-degree-of-freedom (6 DoF) motion platform with a maximum acceleration of 0.5 G, combined with a custom control program connected with the CARLA module of the 3DCoAutoSim simulation environment ~\cite{artal-villa_extension_2019, hussein_3dcoautosim_2018}. The HMI also included a built-in auditory alarm triggered when the inter-platoon distance fell below a predefined threshold. The scenarios featured an automated lead vehicle operating at SAE Level 4 within a highway platooning ODD and a human-driven ego vehicle with SAE Level 3 conditional automation.
This mixed automation aligns with scalable semi-autonomous platooning concepts that accommodate different levels of automation and driver involvement \cite{validi_assessing_2024}.


To investigate whether the HMI supports the driver's supervisory role, an evaluation comprising three driving scenarios in the aforementioned setup was conducted, as detailed in Section IV. The remainder of the paper is structured as follows: In Section II, we review related work on vehicle HMIs. Section III details the design of the developed HMI. 
Section IV describes the experimental design. Section V introduces the data acquisition procedure as well as the defined independent and dependent variables. Section VI presents the evaluation results, and Section VII concludes the paper.

\vspace{-10pt}
\section{Related Work}
In-vehicle HMIs have been extensively investigated in the context of conditional automation and Take Over Requests (TOR). A comprehensive review of TOR characteristics identified interface modality, warning timing, and information content as key factors influencing the safety of control transitions~\cite{morales-alvarez_literature_2020}. 
Experimental studies have examined the impact of different modalities on driver response following a TOR. For example, visual and acoustic warning interfaces were shown to improve driver performance during takeover scenarios when sufficient transition time was available \cite{melcher_take-over_2015}. Similarly, haptic guidance strategies implemented after a TOR were found to influence vehicle control and driving performance during transitions \cite{walter_haptic}. In addition, in-vehicle displays indicating the current driving mode, combined with acoustic warnings, have been shown to support faster and safer driver responses during control transitions \cite{automated-transistions}.


Human factors aspects of Human–Machine Interaction in cooperative and connected vehicle systems have been reviewed in \cite{olaverri-monreal_human_2016}. The study emphasizes the need to assess distraction and workload, apply clear and intuitive warning representations, use appropriate multimodal feedback, ensure consistent visual encoding, such as color coding, and prioritize messages according to their urgency and criticality in V2X-based systems. 

Building on such cooperative V2X communication concepts, visualization approaches have been proposed to extend the driver’s perception beyond line-of-sight obstructions. A VANET-based See-Through System enabled real-time video streaming from a preceding vehicle to support overtaking maneuvers and was evaluated both on-road and in a driving simulator \cite{olaverri-monreal_see-through_2010}. Subsequent work further developed this concept by combining cameras, computer vision, V2V communication, and projection technologies to virtually render preceding vehicles transparent during overtaking \cite{gomez_towards_2026}.

Research has also addressed HMIs in the context of platooning. For example, \cite{gwak_effects_2022} focused on external interfaces.
They evaluated roadside variable message signs in a platoon-merging scenario and rear-mounted LED displays in an emergency stop scenario under varying traffic densities. While these external HMIs showed positive effects such as evasive lane-changing, improved subjective safety, and faster braking responses, in-vehicle HMIs were not evaluated.

With regard to in-vehicle HMIs for platooning, field trials of automated truck platoons have documented their use for monitoring and managing platoon operations, but the HMI specifications have not been introduced, and their effectiveness has not been evaluated \cite{tsugawa_overview_2013}. 
Scalable semi-autonomous platooning frameworks with mixed automation levels and in-vehicle displays have been proposed in \cite{validi_metamodel-based_2022, validi_assessing_2024}. While these works integrate display concepts within system-level evaluations, they do not investigate the behavioral impact of the HMI on driver response and platoon performance.


In another work, \cite{friedrichs_supporting_2016} evaluated in-vehicle HMIs for truck platooning in a driving simulator. Two visual HMIs were compared. The visualizations improved driver trust, reduced crashes, and supported hazard anticipation. However, the mentioned simulation setup did not present a realistic motion platform nor auditory feedback, which significantly reduces the realism of the simulation as well as affecting the final results. 
In the absence of physical motion cues, collisions are reduced to abstract feedback, which may attenuate participants' perceived urgency and thus limit the ecological validity of the observed responses.



A questionnaire-based concept evaluation with truck drivers reported that drivers prefer information about the ego vehicle state and nearby traffic, such as system status, distance, and speed, over information about the overall platoon \cite{10.1145/2851581.2892477}. A driving simulator study confirmed this preference, showing that participants preferred numeric-graphic displays that convey ego-related information \cite{zheng2013human}. A subsequent simulator experiment that included physiological measurements demonstrated that driver stress increased as inter-vehicle distances decreased \cite{zheng2013human, zheng_biosignal_2015}.

A driving simulator study, based on questionnaire evaluations, reported that driver comfort decreased at inter-vehicle distances below approximately 16 meters, while perceived risk increased below approximately 7 meters \cite{larburu2010safe}. The study further indicated that participants considered clear HMI-based information and explicit acknowledgment during transitions between manual and automated platooning to be necessary.

Different levels of automation were also found to influence driver state, with partial automation increasing supervisory workload and higher levels associated with reduced alertness and increased sleepiness \cite{hjalmdahl_driver_2017}. 

Such reductions in vigilance have been discussed in the context of hypovigilance and have motivated the investigation of continuous peripheral visual stimuli in automated driving. In this regard, a tendency towards slightly faster responses to a TOR was observed when drivers were exposed to an unobtrusive luminescent stimulus processed through peripheral vision \cite{capalar_hypovigilance_2017}. 

Such findings underscore the importance of interface designs that balance supervisory demands and mitigate reduced vigilance in conditionally automated driving.


Finally, recent work has proposed new HMIs for truck platooning, such as mirrorless display systems that replace conventional side mirrors with camera-based digital views, combined with bird’s-eye representations of surrounding traffic to support lead-vehicle operation \cite{sugimachi_design_2024}.
In addition, physiological stability indices based on breathing and skin-conductance signals have been introduced to quantify driver anxiety and to evaluate see-through HMIs providing a live forward view from the lead truck, which were shown to enhance driver stability at short gaps \cite{cho_development_2025}. 

To address the gap in the literature regarding driver support in close-interval platooning, we developed an HMI to provide drivers with additional information about the platoon. 
This paper presents custom scenarios implemented in the CARLA module \cite{dosovitskiy_carla_2017} of the 3DCoAutoSim platform using a 6 DoF motion system with a maximum acceleration of 0.5 G. The motion feedback enhances driver engagement, while an acoustic alarm is triggered when the inter-platoon distance falls below a predefined threshold. Compared with static simulators, this setup provides increased realism while preserving the safety benefits of simulation for safety-critical scenarios. 



\begin{figure}[t]
    \centering
    \subfloat[Platooning request (disconnected state). The display shows a textual status indicator, an inactive link symbol representing the uncoupled state, and the current longitudinal distance to the lead vehicle.]{
        \includegraphics[width=0.9\columnwidth]{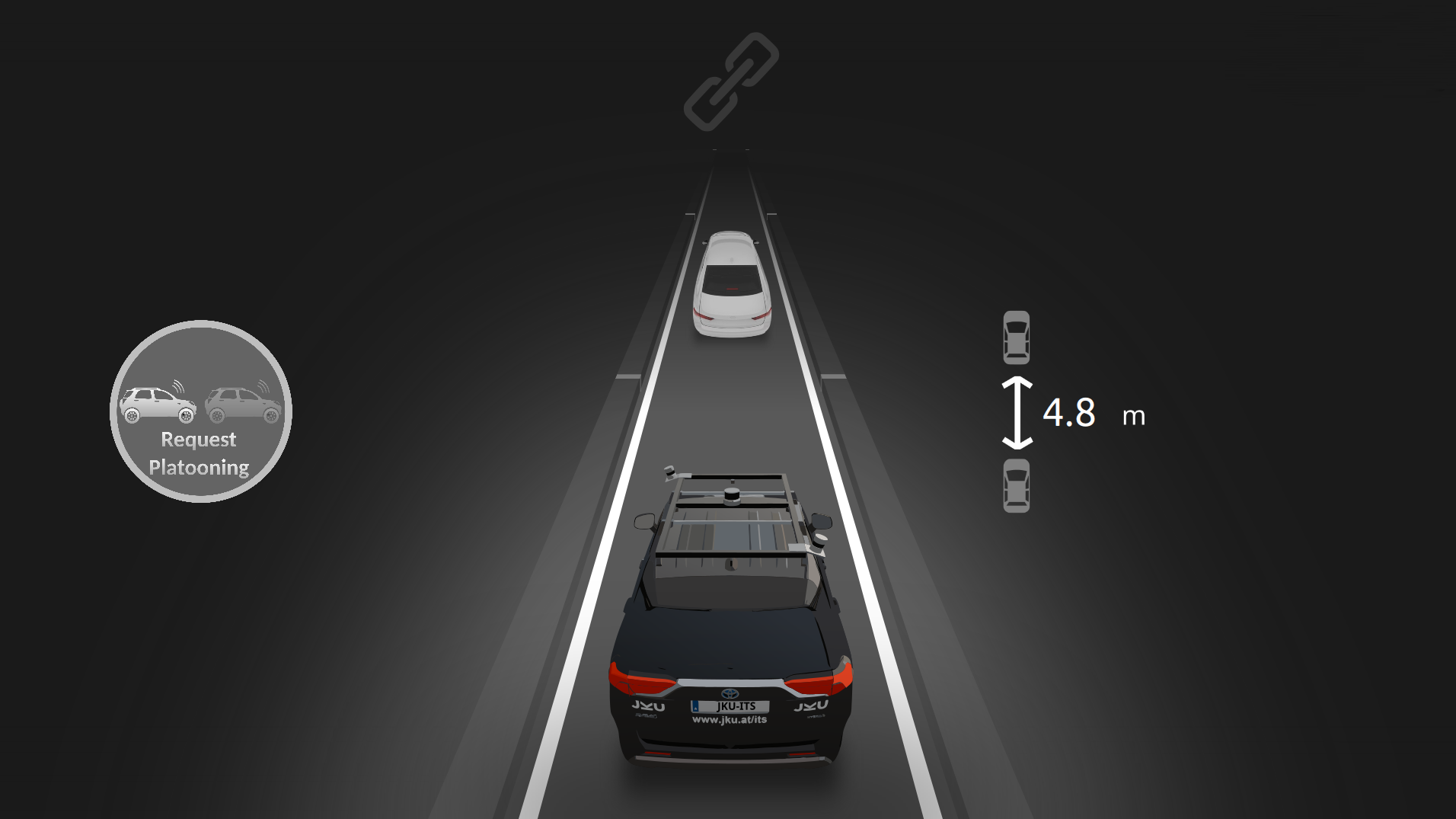}
    }

    \vspace{0.6em}

    \subfloat[Active platooning state (connected state). The textual indicator and link symbol are visually highlighted in green to denote an established coupling, while the numerical distance to the lead vehicle remains continuously displayed.]{
        \includegraphics[width=0.9\columnwidth]{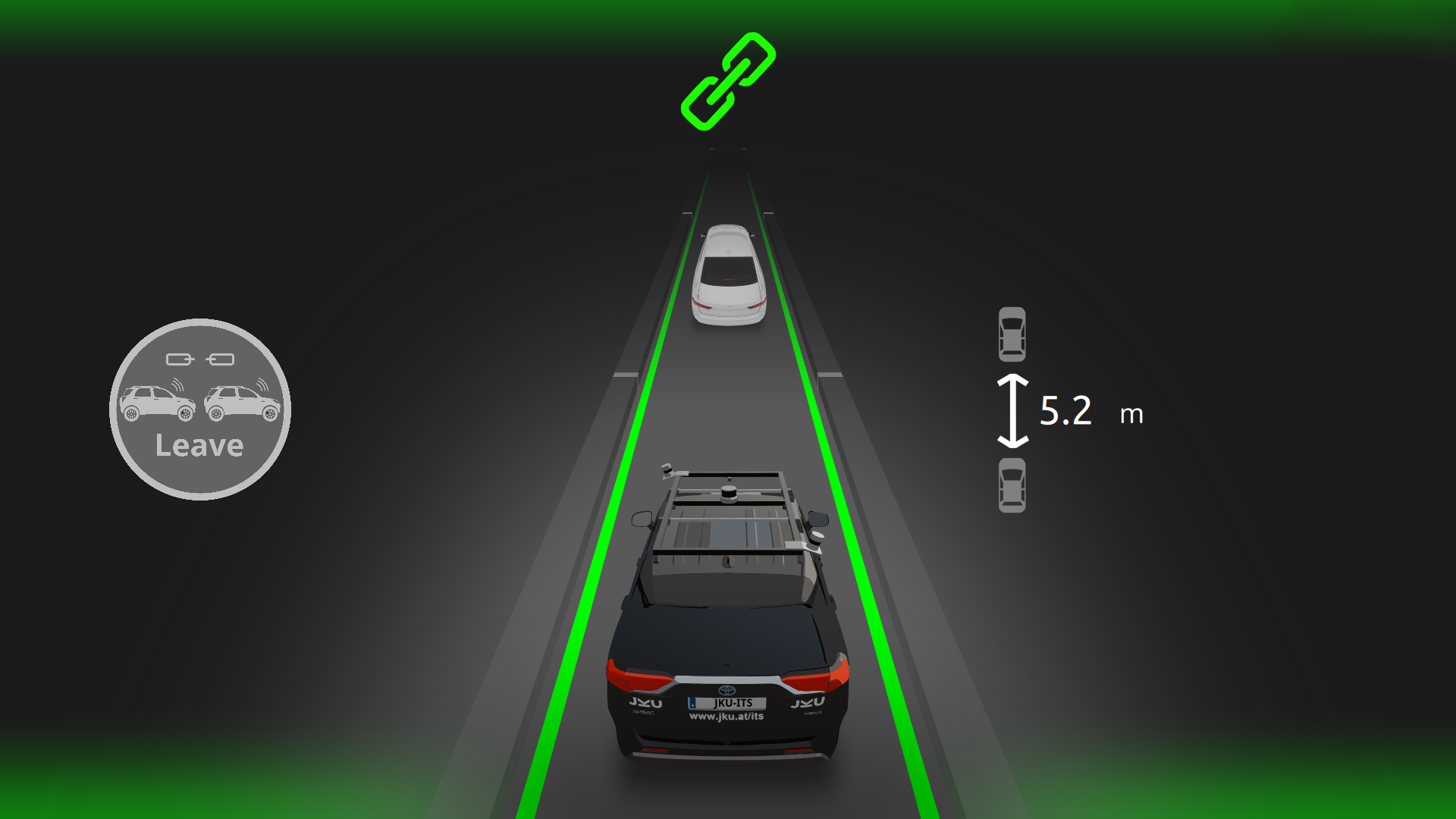}
    }

    \caption{The in-vehicle platooning HMI shows the two system states and their corresponding visual encodings.}
    \label{fig:platooning}
\end{figure}

\vspace{-5pt}
\section{Human-Machine Interface}
The developed HMI consists of three elements: (1) a textual indicator showing whether the platoon is connected or disconnected, (2) a link symbol representing the coupling status between the ego and lead vehicle, and (3) a numerical readout indicating the current longitudinal distance to the lead vehicle in meters.
Figure \ref{fig:platooning}(a) shows the interface in the disconnected state during a platooning request, while Figure \ref{fig:platooning}(b) depicts the active platooning state, in which the textual indicator and link symbol are highlighted in green to signal an established connection.
The HMI was positioned within the driver’s field of view to support continuous situational awareness and enable timely responses to changes in platoon status. 

To evaluate the effect of the HMI on driver behavior and platoon operation, the following hypotheses were formulated:
\begin{description}
    \item[$H_0^1$] The HMI does not improve platoon safety, as measured by the number of collisions.
    \item[$H_0^2$] The HMI does not improve platoon stability, assessed via the number of interventions.
    \item[$H_0^3$] The HMI does not improve driver response, as measured by response latency following platoon disconnection.
\end{description}


\section{Experimental Setup}

\begin{figure}[t]
    \centering
    \subfloat[6 DoF motion platform used for the driving experiment. The in-vehicle HMI is positioned on the right side within the driver’s field of view.]{
        \includegraphics[width=0.6\columnwidth]{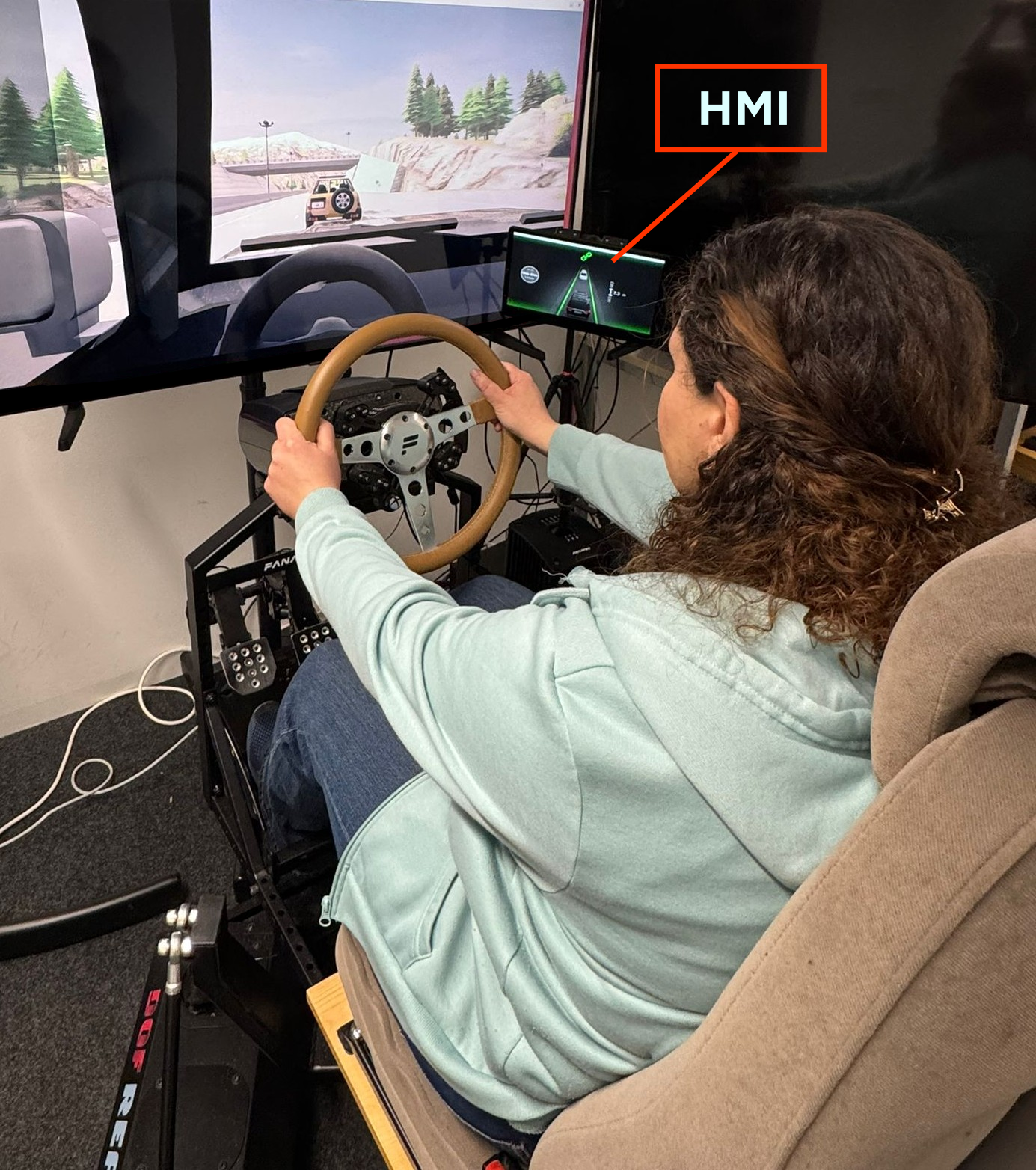}
    }


    \subfloat[The custom platooning scenario showing the automated lead vehicle and the conditionally automated ego vehicle.]{
        \includegraphics[width=0.7\columnwidth]{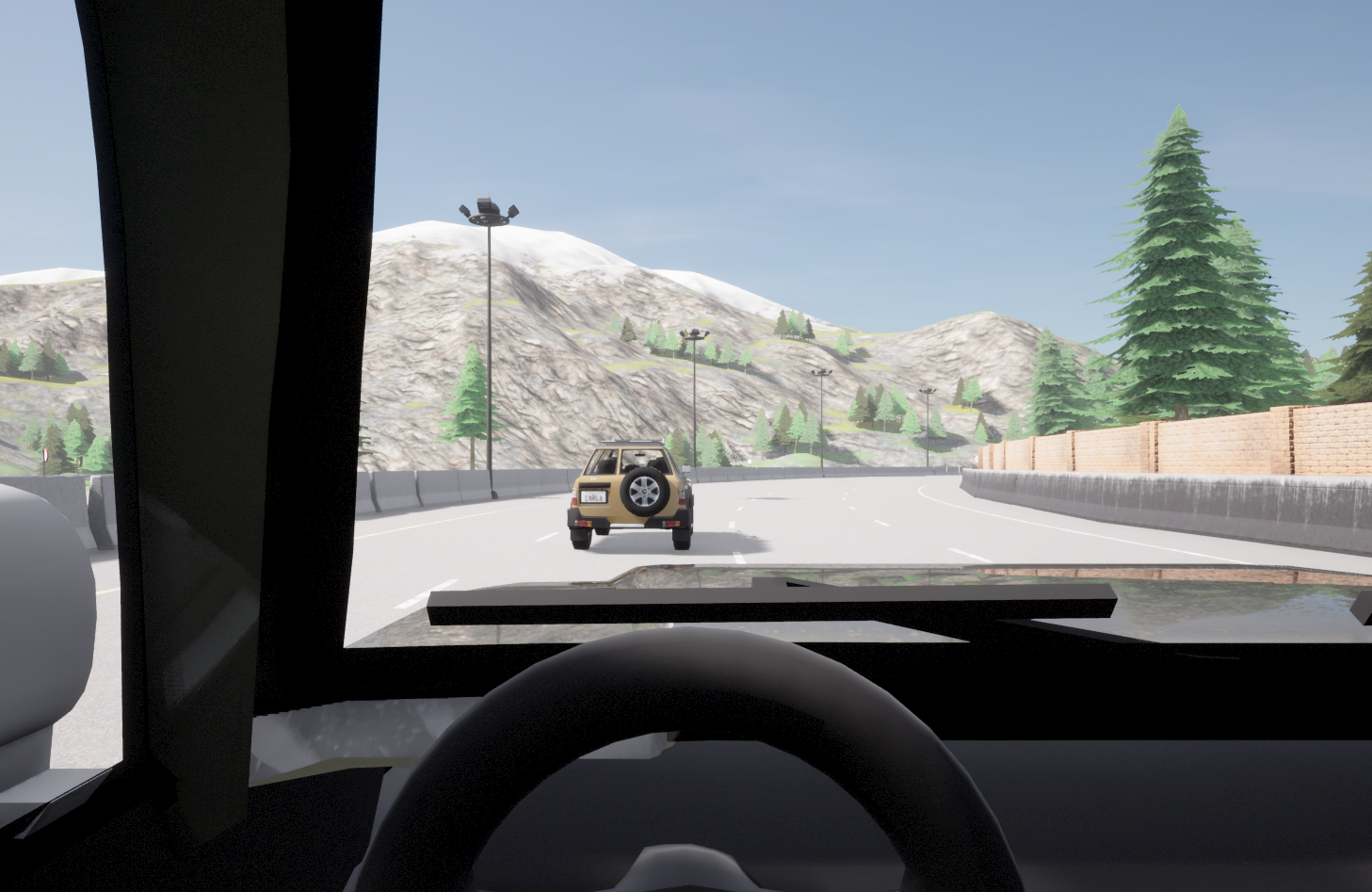}
    }

    \caption{Experimental setup and simulation environment.}
    \label{fig:sim}
\end{figure}

We conducted the experiment using the 3DCoAutoSim platform extended with the CARLA simulation framework. The platform enables the evaluation of cooperative Advanced Driver Assistance Systems (ADAS) through vehicular data exchange among multiple simulators, supporting applications based on Vehicle-to-Vehicle (V2V), Vehicle-to-Infrastructure (V2I), and Vehicle-to-Pedestrian (V2P) communication. It is integrated with the Simulation of Urban Mobility (SUMO) \cite{krajzewicz_recent_2012} for microscopic traffic simulation and congestion analysis \cite{olaverri-monreal_connection_2018}, and with the Robot Operating System (ROS) \cite{quigley_ros_2009} to access software libraries and tools for intelligent vehicle applications \cite{liu2022study}. The platform was executed on an Ubuntu system using a custom-developed control program, coupled to a 6 DoF motion platform, as shown in Figure~\ref{fig:sim} (a).

The scenarios were implemented on a predefined highway-like route consisting of a straight segment, replicating typical motorway platooning conditions.
Each scenario began with the follower vehicle positioned at a fixed distance behind the lead vehicle, with both vehicles assumed to be connected in a platoon configuration from the outset. The lead vehicle initiated acceleration, reaching a target speed of 80 km/h, while the follower vehicle autonomously maintained a 10-meter inter-vehicle gap. Upon reaching a randomized braking point, the lead vehicle applied a predefined braking intensity, classified as mild, moderate, or hard, prompting the follower vehicle to react to avoid a collision.
As the braking intensities were treated as an independent variable, we further describe them in detail in Section~\ref{independentVars}.
Participants were instructed to intervene by using the steering wheel and pedals at their discretion, either to prevent an imminent crash or to adjust the distance to their preferred safety margin.


In all scenarios, the platooning connection was interrupted when a braking event was initiated, allowing participants to take control. 
A braking event was defined as the interval from the point when the lead vehicle brakes until the end of the scenario.
Figure~\ref{fig:sim}~(b) shows a screenshot of the platooning scenario, including the automated lead vehicle and the conditionally automated ego vehicle.

\begin{figure}[t]
    \centering
    \includegraphics[width=1.0\columnwidth]{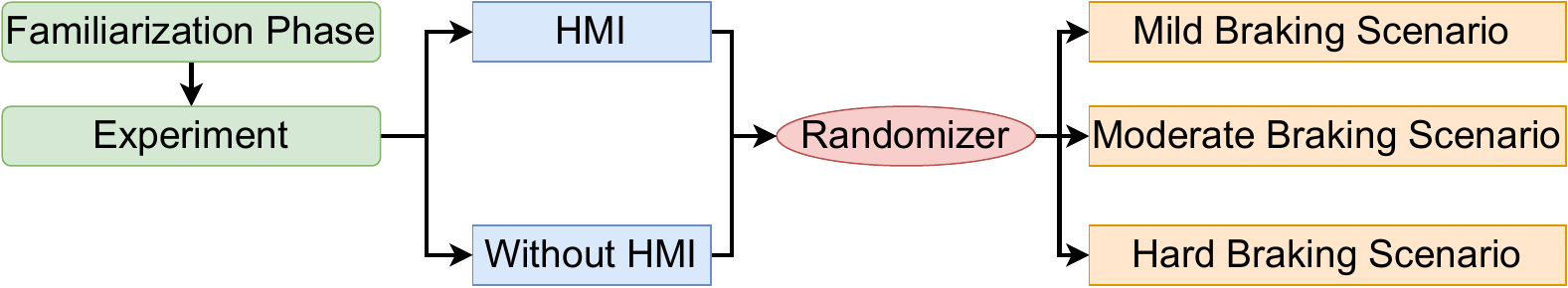}
    \caption{Experimental setup and design overview. Participants first completed a familiarization phase of approximately 5 minutes. The main experimental phase followed a within-subject design, in which each participant completed both HMI conditions. Each condition comprised three braking scenarios (mild, moderate, and hard), presented in a randomized order.}
    \label{fig:setup}
\end{figure}

\section{Data Collection and Analysis}
During the experiments, we logged simulation data, including timestamps, vehicle speed, headway, braking states of the ego and lead vehicle, and the positions of both vehicles. A total of 50 participants took part in the study (mean age = 33.32, SD = 11.10), with a gender distribution of 30 male and 20 female participants. Inclusion criteria were a valid driver’s license and the absence of health conditions that could influence driving behavior, such as color blindness.

Firstly, demographic information was collected. Participants were then provided with standardized instructions regarding the experimental procedure and simulator operation, without being informed about the specific aims or hypotheses of the study.
They were told that they could take over control whenever they deemed necessary and that the HMI did not require manual interaction. 
Participants were also instructed to report any symptoms of motion sickness or discomfort, allowing appropriate measures to be taken promptly. 

Before starting the scenarios, participants completed a phase lasting approximately 5 minutes. During this period, they familiarized themselves with the simulator, as in steering sensitivity, brakes, and throttle, without critical braking events, to be prepared for the main scenarios.

\subsection{Independent Variables} \label{independentVars}

The experiment followed a 2 x 3 within-subject design with two independent variables. 

The HMI condition consisted of two levels:
\begin{itemize}
    \item \textbf{With HMI:} The interface was active.
    \item \textbf{Without HMI:} The interface was disabled.
\end{itemize}

The scenarios were defined by three levels of lead-vehicle braking intensities:
\begin{itemize}
    \item \textbf{Scenario 1 (mild deceleration):} 2.0 m/s\textsuperscript{2}.
    \item \textbf{Scenario 2 (moderate deceleration):} 4.0 m/s\textsuperscript{2}.
    \item \textbf{Scenario 3 (hard deceleration):}  6.0 m/s\textsuperscript{2}.
\end{itemize}

Each participant experienced all combinations of both factors, resulting in six trials per participant, as illustrated in Figure \ref{fig:setup}.

\subsection{Dependent Variables and Metrics}

To objectively evaluate the impact of the HMI, we defined the following dependent variables:

\begin{itemize}


    \item \textbf{Collision occurrence:}  
    A binary variable indicating whether a collision between the ego and lead vehicle occurred during a given trial.


    
    \item \textbf{Response latency:}  
  Response latency was defined as the time interval, measured in seconds, between the start of the braking event and the initiation of braking by the participant. Analyses were restricted to trials in which the ego vehicle maintained a minimum time headway of 2 seconds at the moment the lead vehicle initiated braking, ensuring that only safety-critical events were considered \cite{MAHMUD2017153}.

\item \textbf{Total interventions:}  
Platoon stability was evaluated by counting the total number of manual interventions per scenario prior to the braking event, thereby capturing the phase of potential intact automated platooning. Higher intervention rates were associated with reduced platoon integrity. Participants intervened when they deemed the distance to be too close to the lead vehicle or if they anticipated a crash with the lead vehicle.
\end{itemize}

\subsection{Statistical Analysis}
\vspace{-2pt}
The impact of the HMI condition on each dependent variable was analyzed using mixed-effects models \cite{mcculloch_generalized_2000}. HMI condition and braking scenario were included as fixed factors. 
All analyses were conducted in \texttt{R} \cite{Rsoftware}.

Mixed-effects modeling was selected over traditional approaches such as repeated-measures Analysis of Variance (ANOVA) \cite{ANOVA} because it accommodates non-normal outcome distributions and unequal numbers of observations. 
This was essential given the binary collision outcomes and count data for interventions. 
It allowed the trial order to be included as a covariate to account for temporal effects or condition-order influences.
A total of 300 trials were conducted, of which 293 were included in the analysis. Mixed-effects models incorporated all available observations, allowing participant trials with incomplete data to be used. In contrast, repeated-measures ANOVA typically requires complete data across all conditions, often resulting in the exclusion of participants with missing observations.

For each outcome variable (collision occurrence, total interventions, and response latency), we tested whether the effect of HMI was consistent across participants or varied from person to person. We compared two models: a baseline model that allowed only each participant's average performance to vary (random intercept only), and an extended model that additionally allowed the effect of HMI itself to vary by participant (random intercept and random slope). These two models were tested against each other using a likelihood ratio test (LRT). The more complex model was retained only if the test was significant ($p < .05$) and the model converged without singularity. The random slope was retained only for interventions. Hence, we modeled the other outcomes with a random intercept only. 

Collision occurrence and total interventions were analyzed using generalized linear mixed models (GLMMs) \cite{mcculloch_generalized_2000}, with a binomial distribution specified for collision occurrence.
The total interventions were modeled with a Poisson distribution. In the next step, we conducted Type~III LRTs to evaluate whether each fixed effect contributed significantly to the models and obtained the effect sizes. For each fixed effect, the full models were compared to reduced models in which the respective predictor was omitted, while all other predictors remained in the model. 

The continuous response latency data were analyzed using Linear Mixed Models (LMMs) \cite{mcculloch_generalized_2000}.
Fixed effects were evaluated using Type~III $F$-tests with Satterthwaite-approximated denominator degrees of freedom \cite{satterthwaite}. The Satterthwaite method provides an appropriate approximation of degrees of freedom in mixed-effects models, where exact distributions of test statistics are not available due to random effects.

Finally, to quantify the magnitude of the HMI effect, we reported effect sizes. As different distributions and link functions were used for each dependent variable, the corresponding effect size measures differ. For the logistic collision model, exponentiating the regression coefficient yields an odds ratio \cite{mcculloch_generalized_2000}, representing the multiplicative change in the odds of collision between HMI conditions. For the Poisson intervention model with a log link, exponentiating the coefficient yields a rate ratio, representing the multiplicative change in the expected number of interventions. For the response latency model, exponentiating the estimated contrast yields a geometric mean ratio \cite{GeometricMeanRatio}, representing the multiplicative change in response time between conditions. All effect sizes are reported with 95\% confidence intervals.
\vspace{-2pt}
\section{Results}
\vspace{-2pt}
Table~\ref{tab:generalized} presents the fixed effects for collision occurrence and total interventions. For collisions, neither the HMI condition nor the braking scenario showed a significant effect. However, collision rates were higher with HMI across all scenarios (Scenario~1: 38.8\% vs.\ 20.0\%; Scenario~2: 44.7\% vs.\ 27.1\%; Scenario~3: 32.7\% vs.\ 18.0\%; Table~\ref{tab:collisions}). The corresponding odds ratio (Table~\ref{tab:hmi_effects}) was not significant. 
Accordingly, $H_0^1$ cannot be rejected; the HMI did not significantly affect platoon safety.

\begin{table}[t]
\centering
\caption{Fixed effects from generalized linear mixed models.
LRT are reported for collisions and interventions.}
\begin{tabular}{llccc}
\toprule
Outcome & Effect & $df$ & $\chi^2$ & $p$ \\
\midrule
Collision
& HMI         & 1 & 2.56  & .109 \\
& Scenario    & 2 & 2.60  & .273 \\
& Trial order & 1 & 10.04 & .002$^{**}$ \\
\midrule
Interventions
& HMI         & 1 & 17.72 & $<$.001$^{***}$ \\
& Scenario    & 2 & 0.73  & .695 \\
& Trial order & 1 & 5.30  & .021$^{*}$ \\
\bottomrule
\multicolumn{5}{l}{\footnotesize{$^{*}p < .05$,\ $^{**}p < .01$,\ $^{***}p < .001$}} \\
\end{tabular}
\label{tab:generalized}
\end{table}

In contrast, a statistically significant effect of HMI condition was found for total manual interventions (Table~\ref{tab:generalized}). Participants intervened less frequently when the HMI was active. The rate ratio shown in Table~\ref{tab:hmi_effects} indicates that the expected number of interventions was 80\% higher in the without-HMI condition compared to the with-HMI condition. This pattern was consistent across all three braking scenarios. As presented in Table~\ref{tab:means}, the mean intervention count was lower with the HMI in Scenario~1 (1.18 vs.\ 1.90), Scenario~2 (1.13 vs.\ 1.73), and Scenario~3 (1.06 vs.\ 1.74). Since the difference between interface conditions was statistically significant, $H_0^2$ is rejected. The HMI significantly improved platoon stability, as reflected by reduced manual interventions. From the perspective of the automated system, reduced manual intervention contributes to improved string stability within the platoon. Frequent human overrides can introduce disruptions in the platooning stability. Therefore, the observed reduction in interventions may affect multi-vehicle platoon efficiency and traffic flow.

Regarding response latency, no statistically significant effects of HMI condition, braking scenario, or trial order were observed (Table~\ref{tab:linearmixed}). The geometric mean ratio reported in Table~\ref{tab:hmi_effects} indicates no significant difference between interface conditions. Descriptive values in Table~\ref{tab:means} show minor numerical differences between conditions; however, these differences were not statistically significant. Consequently, $H_0^3$ cannot be rejected. The HMI did not significantly influence driver response latency following platoon disconnection.

Trial order significantly affected collision occurrence ($\chi^2(1)=10.04$, $p=.002$) as well as total interventions ($\chi^2(1)=5.30$, $p=.021$), as shown in Table~\ref{tab:generalized}. 
It was structurally linked to both the HMI condition and the scenario sequence. The observed effects therefore likely reflect not only general adaptation across repeated trials but also adjustments associated with the initial exposure to a specific HMI configuration and the subsequent transition between conditions.

\begin{table}[t]
\centering
\caption{Fixed effects from linear mixed model.
Type III F-tests with Satterthwaite degrees of freedom are reported for response latency.}
\begin{tabular}{llcccc}
\toprule
Outcome & Effect & $df_1$ & $df_2$ & $F$ & $p$ \\
\midrule
Response Latency
& HMI         & 1 & 138.74 & 0.64 & .427 \\
& Scenario    & 2 & 140.15 & 0.19 & .831 \\
& Trial order & 1 & 137.97 & 0.16 & .692 \\
\bottomrule
\end{tabular}
\label{tab:linearmixed}
\end{table}

\begin{table}[t]
\centering
\caption{Exponentiated HMI effects (withoutHMI relative to withHMI).
Ratios are odds ratios (collision), rate ratios (interventions),
and geometric mean ratios (response latency).}
\label{tab:hmi_effects}
\begin{tabular}{lcccc}
\toprule
Outcome & Ratio & Lower CI & Upper CI & $p$ \\
\midrule
Collision        & 0.605 & 0.327 & 1.120 & .108 \\
Interventions    & 1.800 & 1.330 & 2.450 & $<$.001$^{***}$ \\
Response Latency    & 1.137 & 0.829 & 1.559 & .428 \\
\bottomrule
\multicolumn{5}{l}{\footnotesize{$^{*}p < .05$,\ $^{**}p < .01$,\ $^{***}p < .001$}} \\
\end{tabular}
\label{tab:expo}
\end{table}

\begin{table}[t]
\centering
\caption{Mean ($\mu$) and standard deviation ($\sigma$) of response latency and number of interventions per HMI condition and scenario.}
\label{tab:means}
\setlength{\tabcolsep}{4pt}
\begin{tabular}{llcccccc}
\toprule
& & \multicolumn{2}{c}{Scenario 1} & \multicolumn{2}{c}{Scenario 2} & \multicolumn{2}{c}{Scenario 3} \\
\cmidrule(lr){3-4} \cmidrule(lr){5-6} \cmidrule(lr){7-8}
Metric & Condition & $\mu$ & $\sigma$ & $\mu$ & $\sigma$ & $\mu$ & $\sigma$ \\
\midrule
\multirow{2}{*}{Response Latency}
  & With HMI    & 1.07 & 0.62 & 1.48 & 1.71 & 1.57 & 1.68 \\
  & Without HMI & 1.41 & 1.25 & 1.60 & 1.23 & 1.60 & 1.76 \\
\midrule
\multirow{2}{*}{Interventions}
  & With HMI    & 1.18 & 1.54 & 1.13 & 1.33 & 1.06 & 1.09 \\
  & Without HMI & 1.90 & 1.57 & 1.73 & 1.33 & 1.74 & 1.26 \\
\bottomrule
\end{tabular}
\label{tab:Coll}
\end{table}

\begin{table}[t]
\centering
\caption{Number of participants who collided out of total participants
per HMI condition and scenario.}
\label{tab:collisions}
\setlength{\tabcolsep}{5pt}
\begin{tabular}{lcccccc}
\toprule
& \multicolumn{2}{c}{Scenario 1} & \multicolumn{2}{c}{Scenario 2} & \multicolumn{2}{c}{Scenario 3} \\
\cmidrule(lr){2-3} \cmidrule(lr){4-5} \cmidrule(lr){6-7}
Condition & $n$ & \% & $n$ & \% & $n$ & \% \\
\midrule
With HMI    & 19/49 & 38.8 & 21/47 & 44.7 & 16/49 & 32.7 \\
Without HMI & 10/50 & 20.0 & 13/48 & 27.1 &  9/50 & 18.0 \\
\bottomrule
\end{tabular}
\label{tab:Coll}
\end{table}

\section{Conclusion and Future Work}
\vspace{-2pt}
This paper evaluated the impact of an in-vehicle HMI providing continuous system-state and distance information during a platooning scenario coupled with a 6 DoF motion platform and acoustic signals for improved realism. The results demonstrate a significant reduction in manual interventions when the HMI was active. While enhanced transparency generally improved supervisory coordination with the automation, the extent of this improvement varied across drivers. 
The interpretation of these effects should be considered in light of the transition between HMI conditions. Future work will aim to separate interface-specific effects from other influences associated with condition transitions, as well as defining additional factors that may contribute to the observed changes.

From a control perspective, fewer manual overrides imply reduced externally induced disruptions of the longitudinal control loop. In tightly spaced platoons, such reductions are directly relevant to string stability and disturbance propagation, suggesting that informational HMIs can strengthen stability during platooning.
However, while the HMI modified supervisory behavior during platooning, it did not measurably alter reactive braking performance once the platoon was disrupted. 

Future work will investigate further modalities as an extension to the HMI interface to increase perceptual salience in critical situations, while preserving the stability benefits observed during nominal platooning. In addition, evaluating HMIs in longer multi-vehicle platoons and more complex traffic environments will help determine how operational stability can be enhanced without compromising driver safety. Incorporating driver-state measures may further clarify the inter-individual differences observed in supervisory adaptation and support the development of interfaces that balance stability and safety more effectively.

\section*{Acknowledgment}
This work was funded by the Austrian Research Promotion Agency (FFG), PDrive, project number: 12451001.
\vspace{-2pt}

\bibliographystyle{IEEEtran}
\bibliography{final_V2}

@inproceedings{walter_haptic,
  title={Automated driving systems: Impact of haptic guidance on driving performance after a take over request},
  author={Morales-Alvarez, Walter and Certad, Novel and Tadjine, Hadj Hamma and Olaverri-Monreal, Cristina},
  booktitle={2022 IEEE Intelligent Vehicles Symposium (IV)},
  pages={1817--1823},
  year={2022},
  organization={IEEE}
}

@article{satterthwaite,
  title={An approximate distribution of estimates of variance components},
  author={Satterthwaite, Franklin E},
  journal={Biometrics bulletin},
  volume={2},
  number={6},
  pages={110--114},
  year={1946},
  publisher={JSTOR}
}

@book{ANOVA,
  title={Experimental designs using ANOVA},
  author={Tabachnick, Barbara G and Fidell, Linda S},
  volume={724},
  year={2007},
  publisher={Thomson/Brooks/Cole Belmont, CA}
}

@inproceedings{automated-transistions,
	location     = {Changshu},
	title        = {Automated Driving: Interactive Automation Control System to Enhance Situational Awareness in Conditional Automation},
	isbn         = {978-1-5386-4452-2},
	doi          = {10.1109/IVS.2018.8500367},
	shorttitle   = {Automated Driving},
	eventtitle   = {2018 {IEEE} Intelligent Vehicles Symposium ({IV})},
	pages        = {1698--1703},
	booktitle    = {2018 {IEEE} Intelligent Vehicles Symposium ({IV})},
	publisher    = {{IEEE}},
	author       = {Olaverri-Monreal, Cristina and Kumar, Satyarth and Diaz-Alvarez, Alberto},
	date         = {2018-06}
}

@article{sabry2025lidar,
	title        = {A LiDAR-Driven Fallback Longitudinal Controller for Safer Following in Sudden Braking Scenarios},
	author       = {Sabry, Mohamed and Del Re, Enrico and Morales-Alvarez, Walter and Olaverri-Monreal, Cristina},
	journal      = {arXiv preprint arXiv:2509.16642},
	year         = {2025}
}

@inproceedings{artal-villa_extension_2019,
	address      = {Paris, France},
	title        = {Extension of the {3DCoAutoSim} to {Simulate} {Vehicle} and {Pedestrian} {Interaction} based on {SUMO} and {Unity} {3D}},
	copyright    = {https://doi.org/10.15223/policy-029},
	isbn         = {978-1-7281-0560-4},
	doi          = {10.1109/IVS.2019.8814253},
	booktitle    = {2019 {IEEE} {Intelligent} {Vehicles} {Symposium} ({IV})},
	publisher    = {IEEE},
	author       = {Artal-Villa, Leyre and Hussein, Ahmed and Olaverri-Monreal, Cristina},
	month        = jun,
	year         = {2019},
	pages        = {885--890}
}

@inproceedings{gomez_towards_2026,
	title        = {Towards Design Principles for an Accessible Autonomous Vehicle: Promoting Inclusivity, Independence and Well-Being},
	author       = {Gomez, Rafael and Dwyer, James and Peterson, Andrew and Bubke, Alex and Cocks, Kevin and Paz, Alexander},
	booktitle    = {Transport Research Arena Conference},
	pages        = {833--842},
	year         = {2024},
	organization = {Springer}
}

@article{morales-alvarez_literature_2020,
	title        = {Automated {{Driving}}: {{A Literature Review}} of the {{Take}} over {{Request}} in {{Conditional Automation}}},
	shorttitle   = {Automated {{Driving}}},
	author       = {{Morales-Alvarez}, Walter and Sipele, Oscar and L{\'e}beron, R{\'e}gis and Tadjine, Hadj Hamma and {Olaverri-Monreal}, Cristina},
	year         = 2020,
	month        = dec,
	journal      = {Electronics},
	volume       = {9},
	number       = {12},
	publisher    = {publisher},
	issn         = {2079-9292},
	doi          = {10.3390/electronics9122087}
}

@article{olaverri-monreal_human_2016,
	title        = {Human {Factors} in the {Design} of {Human}–{Machine} {Interaction}: {An} {Overview} {Emphasizing} {V2X} {Communication}},
	volume       = {1},
	copyright    = {https://ieeexplore.ieee.org/Xplorehelp/downloads/license-information/IEEE.html},
	issn         = {2379-8904, 2379-8858},
	shorttitle   = {Human {Factors} in the {Design} of {Human}–{Machine} {Interaction}},
	doi          = {10.1109/TIV.2017.2695891},
	number       = {4},
	journal      = {IEEE Transactions on Intelligent Vehicles},
	author       = {Olaverri-Monreal, Cristina and Jizba, Tomas},
	month        = dec,
	year         = {2016},
	pages        = {302--313}
}

@article{GeometricMeanRatio,
	title        = {Ratio of Geometric Means to Analyze Continuous Outcomes in Meta-Analysis: {{Comparison}} to Mean Differences and Ratio of Arithmetic Means Using Empiric Data and Simulation},
	author       = {Friedrich, Jan and Adhikari, Neill and Beyene, Joseph},
	year         = 2012,
	month        = jul,
	journal      = {Statistics in medicine},
	volume       = {31},
	pages        = {1857--86},
	doi          = {10.1002/sim.4501}
}

@book{campolo_vehicular_2015,
	location     = {Cham},
	title        = {Vehicular ad hoc Networks: Standards, Solutions, and Research},
	rights       = {https://www.springernature.com/gp/researchers/text-and-data-mining},
	isbn         = {978-3-319-15496-1 978-3-319-15497-8},
	doi          = {10.1007/978-3-319-15497-8},
	shorttitle   = {Vehicular ad hoc Networks},
	publisher    = {Springer International Publishing},
	editor       = {Campolo, Claudia and Molinaro, Antonella and Scopigno, Riccardo},
	date         = {2015}
}

@book{ulsoy_automotive_2012,
	location     = {New York},
	title        = {Automotive Control Systems},
	isbn         = {978-1-107-01011-6 978-1-139-41973-4},
	pagetotal    = {1},
	publisher    = {Cambridge University Press},
	author       = {Ulsoy, A. Galip},
	editora      = {Peng, Huei and Çakmakci, Melih},
	editoratype  = {collaborator},
	date         = {2012}
}

@article{sarter_how_1995,
	title        = {How in the World Did We Ever Get into That Mode? Mode Error and Awareness in Supervisory Control},
	volume       = {37},
	issn         = {0018-7208, 1547-8181},
	doi          = {10.1518/001872095779049516},
	pages        = {5--19},
	number       = {1},
	journaltitle = {Human Factors: The Journal of the Human Factors and Ergonomics Society},
	author       = {Sarter, Nadine B. and Woods, David D.},
	date         = {1995},
	langid       = {english}
}

@article{jamieson_designing_2005,
	title        = {Designing Effective Human-Automation-Plant Interfaces: A Control-Theoretic Perspective},
	volume       = {47},
	issn         = {0018-7208, 1547-8181},
	doi          = {10.1518/0018720053653820},
	shorttitle   = {Designing Effective Human-Automation-Plant Interfaces},
	pages        = {12--34},
	number       = {1},
	journaltitle = {Human Factors: The Journal of the Human Factors and Ergonomics Society},
	shortjournal = {Hum Factors},
	author       = {Jamieson, Greg A. and Vicente, Kim J.},
	date         = 2005
}

@article{validi_assessing_2024,
	title        = {Assessing energy consumption in scalable semi-autonomous destination-based {E}-platoons: {A} multiplayer approach},
	volume       = {136},
	issn         = {1361-9209},
	shorttitle   = {Assessing energy consumption in scalable semi-autonomous destination-based {E}-platoons},
	doi          = {10.1016/j.trd.2024.104464},
	journal      = {Transportation Research Part D: Transport and Environment},
	author       = {Validi, Aso and Liu, Yuzhou and Olaverri-Monreal, Cristina},
	month        = nov,
	year         = 2024,
	pages        = {104464}
}

@article{melcher_take-over_2015,
	title        = {Take-Over Requests for Automated Driving},
	volume       = {3},
	issn         = {23519789},
	doi          = {10.1016/j.promfg.2015.07.788},
	pages        = {2867--2873},
	journaltitle = {Procedia Manufacturing},
	shortjournal = {Procedia Manufacturing},
	author       = {Melcher, Vivien and Rauh, Stefan and Diederichs, Frederik and Widlroither, Harald and Bauer, Wilhelm},
	date         = {2015},
	langid       = {english}
}

@inproceedings{olaverri-monreal_see-through_2010,
	address      = {La Jolla, CA, USA},
	title        = {The {See}-{Through} {System}: {A} {VANET}-enabled assistant for overtaking maneuvers},
	isbn         = {978-1-4244-7866-8},
	shorttitle   = {The {See}-{Through} {System}},
	doi          = {10.1109/IVS.2010.5548020},
	booktitle    = {2010 {IEEE} {Intelligent} {Vehicles} {Symposium}},
	publisher    = {IEEE},
	author       = {Olaverri-Monreal, Cristina and Gomes, Pedro and Fernandes, Ricardo and Vieira, Fausto and Ferreira, Michel},
	month        = jun,
	year         = {2010},
	pages        = {123--128}
}

@article{gwak_effects_2022,
	title        = {Effects of Traffic-related Environmental Factors on Acceptability and Safety of Truck Platooning for Peripheral Drivers: A Simulator Study},
	volume       = {13},
	issn         = {2185-0984, 2185-0992},
	doi          = {10.20485/jsaeijae.13.2_54},
	shorttitle   = {Effects of Traffic-related Environmental Factors on Acceptability and Safety of Truck Platooning for Peripheral Drivers},
	pages        = {54--59},
	number       = {2},
	journaltitle = {International Journal of Automotive Engineering},
	shortjournal = {{IJAE}},
	author       = {Gwak, Jongseong and Shimono, Keisuke and Suda, Yoshihiro},
	date         = {2022},
	langid       = {english}
}

@article{tsugawa_overview_2013,
	title        = {An Overview on an Automated Truck Platoon within the Energy {ITS} Project},
	volume       = {46},
	issn         = {14746670},
	doi          = {10.3182/20130904-4-JP-2042.00110},
	pages        = {41--46},
	number       = {21},
	journaltitle = {{IFAC} Proceedings Volumes},
	shortjournal = {{IFAC} Proceedings Volumes},
	author       = {Tsugawa, Sadayuki},
	date         = {2013},
	langid       = {english}
}

@article{validi_metamodel-based_2022,
	title        = {Metamodel-based simulation to assess platooning on battery energy consumption},
	volume       = {109},
	issn         = {13619209},
	doi          = {10.1016/j.trd.2022.103350},
	language     = {en},
	journal      = {Transportation Research Part D: Transport and Environment},
	author       = {Validi, Aso and Smirnov, Nikita and Olaverri-Monreal, Cristina},
	month        = aug,
	year         = {2022},
	pages        = {103350}
}

@inproceedings{friedrichs_supporting_2016,
	location     = {Ann Arbor {MI} {USA}},
	title        = {Supporting Drivers in Truck Platooning: Development and Evaluation of Two Novel Human-Machine Interfaces},
	isbn         = {978-1-4503-4533-0},
	doi          = {10.1145/3003715.3005451},
	shorttitle   = {Supporting Drivers in Truck Platooning},
	eventtitle   = {{AutomotiveUI}'16: 8th International Conference on Automotive User Interfaces and Interactive Vehicular Applications},
	pages        = {277--284},
	booktitle    = {Proceedings of the 8th International Conference on Automotive User Interfaces and Interactive Vehicular Applications},
	publisher    = {{ACM}},
	author       = {Friedrichs, Thomas and Ostendorp, Marie-Christin and Lüdtke, Andreas},
	date         = {2016-10-24},
	langid       = {english}
}

@inproceedings{10.1145/2851581.2892477,
	location     = {San Jose, California, {USA}},
	title        = {Design of a human-machine interface for truck platooning},
	isbn         = {978-1-4503-4082-3},
	doi          = {10.1145/2851581.2892477},
	series       = {Chi ea '16},
	pages        = {2285--2291},
	booktitle    = {Proceedings of the 2016 {CHI} conference extended abstracts on human factors in computing systems},
	publisher    = {Association for Computing Machinery},
	author       = {Sadeghian Borojeni, Shadan and Friedrichs, Thomas and Heuten, Wilko and Lüdtke, Andreas and Boll, Susanne},
	date         = {2016}
}

@inproceedings{zheng2013human,
	title        = {Human-machine interface system for simulation-based automatic platooning of trucks},
	pages        = {535--539},
	booktitle    = {16th international {IEEE} conference on intelligent transportation systems ({ITSC} 2013)},
	publisher    = {{IEEE}},
	author       = {Zheng, Rencheng and Nakano, Kimihiko and Kato, Shin and Ogitsu, Takeki and Yamabe, Shigeyuki and Aoki, Keiji and Suda, Yoshihiro},
	date         = {2013}
}

@article{zheng_biosignal_2015,
	title        = {Biosignal Analysis to Assess Mental Stress in Automatic Driving of Trucks: Palmar Perspiration and Masseter Electromyography},
	volume       = {15},
	rights       = {http://creativecommons.org/licenses/by/3.0/},
	issn         = {1424-8220},
	doi          = {10.3390/s150305136},
	shorttitle   = {Biosignal Analysis to Assess Mental Stress in Automatic Driving of Trucks},
	pages        = {5136--5150},
	number       = {3},
	journaltitle = {Sensors},
	publisher    = {Multidisciplinary Digital Publishing Institute},
	author       = {Zheng, Rencheng and Yamabe, Shigeyuki and Nakano, Kimihiko and Suda, Yoshihiro},
	date         = {2015-03}
}

@inproceedings{larburu2010safe,
	title        = {Safe road trains for environment: Human factors’ aspects in dual mode transport systems},
	pages        = {1--12},
	booktitle    = {{ITS} world congress, busan, korea},
	author       = {Larburu, Maider and Sanchez, Javier and Rodriguez, Domingo José},
	date         = 2010
}

@article{hjalmdahl_driver_2017,
	title        = {Driver behaviour and driver experience of partial and fully automated truck platooning – a simulator study},
	volume       = {9},
	issn         = {1866-8887},
	doi          = {10.1007/s12544-017-0222-3},
	pages        = {8},
	number       = {1},
	journaltitle = {European Transport Research Review},
	shortjournal = {Eur. Transp. Res. Rev.},
	author       = {Hjälmdahl, Magnus and Krupenia, Stas and Thorslund, Birgitta},
	date         = {2017-02-02},
	langid       = {english}
}

@inproceedings{capalar_hypovigilance_2017,
	title        = {Hypovigilance in limited self-driving automation: {Peripheral} visual stimulus for a balanced level of automation and cognitive workload},
	issn         = {2153-0017},
	shorttitle   = {Hypovigilance in limited self-driving automation},
	doi          = {10.1109/ITSC.2017.8317925},
	booktitle    = {2017 {IEEE} 20th {International} {Conference} on {Intelligent} {Transportation} {Systems} ({ITSC})},
	author       = {Çapalar, Jusuf and Olaverri-Monreal, Cristina},
	month        = oct,
	year         = {2017},
	pages        = {27--31}
}

@inproceedings{sugimachi_design_2024,
	title        = {Design of Human-Machine Interface for Truck Platooning Using Driving Simulator},
	volume       = {159},
	isbn         = {978-1-964867-35-9},
	issn         = {27710718},
	doi          = {10.54941/ahfe1005676},
	eventtitle   = {{AHFE} (2024) International Conference},
	number       = {159},
	booktitle    = {Human Factors in Design, Engineering, and Computing},
	publisher    = {{AHFE} Open Acces},
	author       = {Sugimachi, Toshiyuki},
	date         = {2024}
}

@article{cho_development_2025,
	title        = {Development of a Stability Index for Evaluating Drivers’ Psychological Stability During Truck Platooning},
	volume       = {15},
	rights       = {http://creativecommons.org/licenses/by/3.0/},
	issn         = {2076-3417},
	doi          = {10.3390/app15105429},
	pages        = {5429},
	number       = {10},
	journaltitle = {Applied Sciences},
	publisher    = {Multidisciplinary Digital Publishing Institute},
	author       = {Cho, Hyonbae and Kim, Yejin and Oh, {SeokJin} and Yun, Ilsoo},
	date         = {2025-01},
	langid       = {english}
}

@inproceedings{hussein_3dcoautosim_2018,
	address      = {Maui, HI},
	title        = {{3DCoAutoSim}: {Simulator} for {Cooperative} {ADAS} and {Automated} {Vehicles}},
	isbn         = {978-1-7281-0321-1 978-1-7281-0323-5},
	shorttitle   = {{3DCoAutoSim}},
	doi          = {10.1109/ITSC.2018.8569512},
	booktitle    = {2018 21st {International} {Conference} on {Intelligent} {Transportation} {Systems} ({ITSC})},
	publisher    = {IEEE},
	author       = {Hussein, Ahmed and Diaz-Alvarez, Alberto and Armingol, Jose Maria and Olaverri-Monreal, Cristina},
	month        = nov,
	year         = {2018},
	pages        = {3014--3019}
}

@misc{dosovitskiy_carla_2017,
	title        = {{CARLA}: An Open Urban Driving Simulator},
	doi          = {10.48550/arXiv.1711.03938},
	shorttitle   = {{CARLA}},
	number       = {{arXiv}:1711.03938},
	publisher    = {{arXiv}},
	author       = {Dosovitskiy, Alexey and Ros, German and Codevilla, Felipe and Lopez, Antonio and Koltun, Vladlen},
	date         = {2017-11-10},
	eprinttype   = {arxiv},
	eprint       = {1711.03938 [cs]}
}

@article{krajzewicz_recent_2012,
	title        = {Recent development and applications of {SUMO}-{Simulation} of {Urban} {MObility}},
	volume       = {5},
	number       = {3\&4},
	journal      = {International journal on advances in systems and measurements},
	author       = {Krajzewicz, Daniel and Erdmann, Jakob and Behrisch, Michael and Bieker, Laura and {others}},
	year         = {2012},
	pages        = {128--138}
}

@article{olaverri-monreal_connection_2018,
	title        = {Connection of the {SUMO} {Microscopic} {Traffic} {Simulator} and the {Unity} {3D} {Game} {Engine} to {Evaluate} {V2X} {Communication}-{Based} {Systems}},
	volume       = {18},
	issn         = {1424-8220},
	doi          = {10.3390/s18124399},
	language     = {en},
	number       = {12},
	journal      = {Sensors},
	author       = {Olaverri-Monreal, Cristina and Errea-Moreno, Javier and Díaz-Álvarez, Alberto and Biurrun-Quel, Carlos and Serrano-Arriezu, Luis and Kuba, Markus},
	month        = dec,
	year         = {2018},
	pages        = {4399}
}

@inproceedings{quigley_ros_2009,
	title        = {{ROS}: an open-source {Robot} {Operating} {System}},
	volume       = {3},
	number       = {3.2},
	booktitle    = {{ICRA} workshop on open source software},
	publisher    = {Kobe},
	author       = {Quigley, Morgan and Conley, Ken and Gerkey, Brian and Faust, Josh and Foote, Tully and Leibs, Jeremy and Wheeler, Rob and Ng, Andrew Y and {others}},
	year         = {2009},
	pages        = {5}
}

@inproceedings{liu2022study,
	title        = {Study of {ROS}-based autonomous vehicles in snow-covered roads by means of behavioral cloning using {3DCoAutoSim}},
	booktitle    = {World conference on information systems and technologies},
	publisher    = {Springer},
	author       = {Liu, Yuzhou and Morales-Alvarez, Walter and Novotny, Georg and Olaverri-Monreal, Cristina},
	year         = {2022},
	pages        = {211--221}
}

@book{mcculloch_generalized_2000,
	edition      = {1},
	series       = {Wiley {Series} in {Probability} and {Statistics}},
	title        = {Generalized, {Linear}, and {Mixed} {Models}},
	copyright    = {http://doi.wiley.com/10.1002/tdm\_license\_1.1},
	isbn         = {978-0-471-19364-7 978-0-471-72207-6},
	doi          = {10.1002/0471722073},
	language     = {en},
	publisher    = {Wiley},
	author       = {McCulloch, Charles E. and Searle, Shayle R.},
	month        = dec,
	year         = {2000}
}

@article{MAHMUD2017153,
	title        = {Application of Proximal Surrogate Indicators for Safety Evaluation: {{A}} Review of Recent Developments and Research Needs},
	author       = {Mahmud, S.M. Sohel and Ferreira, Luis and Hoque, Md. Shamsul and Tavassoli, Ahmad},
	year         = 2017,
	journal      = {IATSS Research},
	volume       = {41},
	number       = {4},
	pages        = {153--163},
	issn         = {0386-1112},
	doi          = {10.1016/j.iatssr.2017.02.001}
}

@Manual{Rsoftware,
	title        = {R: A Language and Environment for Statistical Computing},
	author       = {{R Core Team}},
	organization = {R Foundation for Statistical Computing},
	address      = {Vienna, Austria},
	year         = {2021}
}

@article{van2006impact,
	title        = {The impact of cooperative adaptive cruise control on traffic-flow characteristics},
	author       = {Van Arem, Bart and Van Driel, Cornelie JG and Visser, Ruben},
	journal      = {IEEE Transactions on intelligent transportation systems},
	volume       = {7},
	number       = {4},
	pages        = {429--436},
	year         = {2006},
	publisher    = {IEEE}
}
\end{document}